# Schottky diode temperature sensor for pressure sensor

**M. Basov ***
Dukhov Automatics Research Institute (VNIIA), Moscow 127055, Russian Federation
*Corresponding author. *E-mail address:* engineerbasovm@gmail.com (M. Basov)

ABSTRACT

The small silicon chip of Schottky diode (0.8x0.8x0.4 $mm^3$) with planar arrangement of electrodes (chip PSD) as temperature sensor, which functions under the operating conditions of pressure sensor, was developed. The forward I-V characteristic of chip PSD is determined by potential barrier between Mo and n-Si ($N_D = 3 \times 10^{15}$ $cm^{-3}$). Forward voltage $U_F = 208 \pm 6$ mV and temperature coefficient TC = $-1.635 \pm 0.015$ mV/ºC (with linearity $k_T$ <0.4% for temperature range of -65 to +85 ºC) at supply current $I_F = 1$ mA is achieved. The reverse I-V characteristic has high breakdown voltage $U_{BR} > 85$ V and low leakage current $I_L < 5$ μA at 25 ºC and $I_L < 130$ μA at 85 ºC ($U_R = 20$ V) because chip PSD contains the structure of two p-type guard rings along the anode perimeter. The application of PSD chip for wider temperature range from -65 to +115 ºC is proved. The separate chip PSD of temperature sensor located at a distance of less than 1.5 mm from the pressure sensor chip. The PSD chip transmits input data for temperature compensation of pressure sensor errors by ASIC and for direct temperature measurement.

Keywords: temperature sensor, Schottky diode, Mo/Si-n barrier, guard rings, pressure sensor.

## 1. Introduction

Semiconductor temperature sensors are among the most applicable elements for analysis of physical and chemical processes in a wide range of industries and research. Today temperature sensors are used in automotive, medical, aviation, space, nuclear, metallurgical and many other industries [1-8]. These elements find solutions in IoT and customers application (smart devices and smart home, HVAC, AI mechanisms) [9, 10], are created together with microprocessors and electronic devices using CMOS technologies [1, 10-13], measure temperature for scientific research (biophysics and biochemistry, robotics, metrology, etc.) [2, 14, 15]. The conditions are determined by the field of application with following factors: temperature range, temperature coefficient, measurement error, overall dimensions or size, power consumption [13, 16-19], state of aggregation and aggressiveness of measured environment (radiation, high temperatures and corrosion) [11, 16, 17, 20, 21]. Temperature sensors are created by various types of semiconductor structures: thermistors [1-3, 22, 23], p-n junction diodes [12, 24, 25], JFET [1], fiber optic [26] and capacitive sensors [4, 6]. Schottky diode (SD) temperature sensors should be noted separately, because their constructions are extensive. SDs use epitaxial silicon [19, 25, 27-29] and polysilicon structures [11] of n - or p-type conductivity with array of different metal, silicon carbide 4H-SiC with barrier metal of Ni [16, 20], Ti [30, 31], Pt [32] and $V_2O_5$ [18], as well as working layers of GaN [17, 33] or graphene [34, 35] and many other combinations. Semiconductor structures and sapphire [17] or diamond [19] insulators were chosen as substrate for developments. Additionally, SDs were created on SOI wafers [12, 13, 34, 36]. SD temperature sensors were shown as separate components or within various electrical connections, for example: CTAT and PTAT [10] or Wheatstone bridge circuits [3]. Today one of the relevant directions for sensor development is the measurement of several properties for environment in a single device [37]. For example, there are methods for simultaneous measurement of temperature and pressure by a single element [4, 6, 26] or various elements within a single chip [12, 23]. The use of similar methods for one chip or individual chips in a single case is determined by operating conditions, but on the other hand, by the ratio between: 1) chip price, 2) choice of wafer structure and capabilities of fabrication technology, 3) complexity of technological route, especially in the conditions of combining between CMOS and MEMS processes, 4) yield of each element, 5) design features of assembly [38].

Based on all conditions above, the structure of separate temperature sensor chip in the form of SD, developed for joint use with pressure sensor, was chosen. SD temperature sensor was formed by planar technology with arrangement of anode and cathode on the same wafer surface (PSD chip). Further, PSD chips were placed in a single case at a distance of less than 1.5 mm from the pressure sensor chip. The creation of temperature sensor as separate chip allows elements to be independent from each other regarding the choice of initial semiconductor material, combination of technological processes and, most importantly, methods of pressure and temperature measurement. Pressure sensor can operate on the piezoresistive effect, using single sensitive element [39], classical Wheatstone bridge electrical circuit [40-45] or new development utilizing piezosensitive differential amplifier with negative feedback loop (PDA-NFL) circuit [46-50], or any other effects [4, 6, 23, 26, 37]. An additional advantage of temperature sensor creating as a separate chip is no effect of residual mechanical stresses from applying pressure to pressure sensor membrane. The mechanical stress could significantly influence the current-voltage (I–V) characteristics of a temperature sensor [51-53].

## 2. Development of PSD chip

The reasons for using SD temperature sensor are determined by its properties, which guarantee the following achievement of operational needs:

- Low forward voltage $U_F$ at supply current $I_F = 1$ mA, which declared by ASIC for pressure sensor.
- Low leakage currents ($I_L < 10$ µA at $U_R = 20$ V) and high breakdown voltage ($V_{BD} > 80$ V) at $T_{room}$ on reverse branch of I–V characteristic required for linearity of temperature coefficient (TC) at elevated temperatures [12, 13, 54] and combining SD cathode with "ground" contact of pressure sensor circuit.
- High TC values ($|TC| > 1.6$ mV/ºC) with low linearity error (dTC < 0.5%) in temperature range from -65 to 85 ºC.
- Small dimensions of chip for use with pressure sensor chip in a single case or other small-sized devices for future developments.

As it is known, the physical principle of SD operation is based on a barrier potential difference, for example, between semiconductor of n-type conductivity with low concentration and metal with a large work function [54, 55], which blocked the free emission of electrons from the metal. SDs can be created on both n-type and p-type conductivity semiconductors, but n-type conductivity is preferable due to the higher electron mobility [56]. The current through the SD thus obtained was analysed in terms of the thermionic emission diffusion equation or Richardson equation [17, 18, 34, 57-59]:

$$I = AA^*T^2 \exp(-\frac{q\varphi_{B0}}{kT}) \cdot \exp(\frac{qV}{nkT}) \cdot (1 - \exp(\frac{qV}{kT})), \quad (1)$$

where A – anode area, A* – Richardson constant equal to 112 $A/cm^2/K^2$ for n-type conductivity and 32 $A/cm^2/K^2$ for p-type conductivity, T – temperature, q – electron charge equal to $1{,}6 \cdot 10^{-19}$ C, $\varphi_{B0}$ – Schottky barrier height, k – Boltzmann constant equal to $8{,}62 \cdot 10^{-6}$ eV/K, n – ideality factor, V – external voltage. It should be noted that I-V characteristic of forward bias is significantly affected by the surface state of semiconductor, which is determined by the purity of preparation before the barrier metal deposition in production process. TC for forward bias voltage of SD in the open mode (like the current) depends on ideality factor n and barrier height $\varphi_{B0}$, which is determined by the choice of metal and concentration of carriers in semiconductor $N_D$ [27–29, 36]:

$$\frac{dU}{dT} = (\frac{d\varphi_{B0}}{dT}) + (nk \ln(\frac{I}{AA^*T^2}) - 2). \quad (2)$$

The current for reverse branch of I–V characteristic can be calculated [44, 45, 52, 56]:

$$I_R = AA^*T^2 \exp(-\frac{q\varphi_{B0}}{kT}) \cdot \exp(\frac{q\sqrt{qE/4\pi\varepsilon_S}}{kT}), \quad (3)$$

where $\varepsilon_S$ is the permittivity (dielectric constant) equal to $11.8 \cdot \varepsilon_0 = 1.04 \cdot 10^{-12}$ F/cm and the value of E is defined as:

$$E = \sqrt{\frac{2qN_D}{\varepsilon_S}(V + V_{bi} - \frac{kT}{q})}, \quad (4)$$

where $N_D$ – concentration of carriers in n-type semiconductor, $V_{bi}$ – potential of built-in charge. The reason for low SD breakdown voltage is the edge leakage currents along the surface. High doped $p^+$-type region, which is formed of guard ring (GR) along the perimeter of the contact window for SD anode, can reduce electric field in Schottky barrier area and as a result increase the breakdown voltage [31, 60-63]. There are known methods for redistribution of charge carriers, when the space charge regions (SCR) of two or more neighboring GRs intersect between each other at a time close to their single avalanche breakdown at p-n junction. It helps to increase breakdown voltage of SD relative to structure with one GR. The increase of breakdown voltage reduces sharp growth of leakage current by temperature increasing and, therefore, expand a temperature range of PSD chip while maintaining low errors.

The PSD chip was created on epitaxial silicon wafers (epitaxial layer of n-type conductivity (100): $W_{ep}$ = 15 µm, $\rho_{ep} \approx 2.2$ Ohm·cm, $N_{D\,ep} = 3 \cdot 10^{15}$ $cm^{-3}$; substrate $n^+$-type of conductivity (100): $W_{sub}$ = 380 µm, $\rho_{sub} \approx 0.01$ Ohm·cm). The overall dimensions of PSD chip are 0.8x0.8x0.4 $mm^3$ (anode area is 0.24 $mm^2$), which are determined by conditions of chip location on a case pin with pressure sensor and application with $I_{sup/F}$ = 1 mA. One wafer with diameter of 3 inches has more than 1150 simples of PSD chip. The barrier metals for development are Al and Mo layers. The structures of PSD chip with one or two GRs were used for analysis of breakdown voltage increasing. The technology route has followed process sequence (relevant photolithography (PL) steps):

- Creation of deep high-doped region with $n^+$-type conductivity ($x_{n+}$ = 7,6 µm, $R_{S\,p+}$ = 1,2 Ohm/$cm^2$, $N_{S\,p+} = 3 \cdot 10^{18}$ $cm^{-3}$) in epitaxial layer to form a ohmic contact to substrate. The equipotential location of this region on the front wafer side creates possibility to form a SD cathode and anode on single surface.

- Formation of high-doped region with $p^+$-type conductivity ($x_{p+}$ = 1.3 μm, $R_{Sp+}$ = 71 Ohm/cm$^2$) as GRs. The development uses the creation of two types of PSD chip, which is implemented on the one plate: the PSD chip No. 1 has one GR, which is located along a perimeter of anode contact window; the PSD chip No. 2 has two GRs, where the additional structure of RD has a wider radius.
- Etching of $SiO_2$ for contact windows in the areas where SD electrodes are located.
- Deposition of Al or Mo barrier layers ($W_{Mo}$ = 0.2 μm) on the anode area.

The final deposition process of relatively thicker Al layer ($W_{Al}$ = 0.8 μm) for two SD electrode areas. It is necessary for application of ultrasonic welding for chip contact pads. The previous iteration (PL step) is accordingly unnecessary in case of using Al as a barrier metal.

## 3. Experimental research of PSD chip

Fig. 1 shows images of PSD chip variants, schematic section of structure with two GRs and Mo barrier metal. Fig. 1(d) presents a method of PSD chip landing in a single case with a pressure sensor at the distance of less than 1.5 mm from each other.

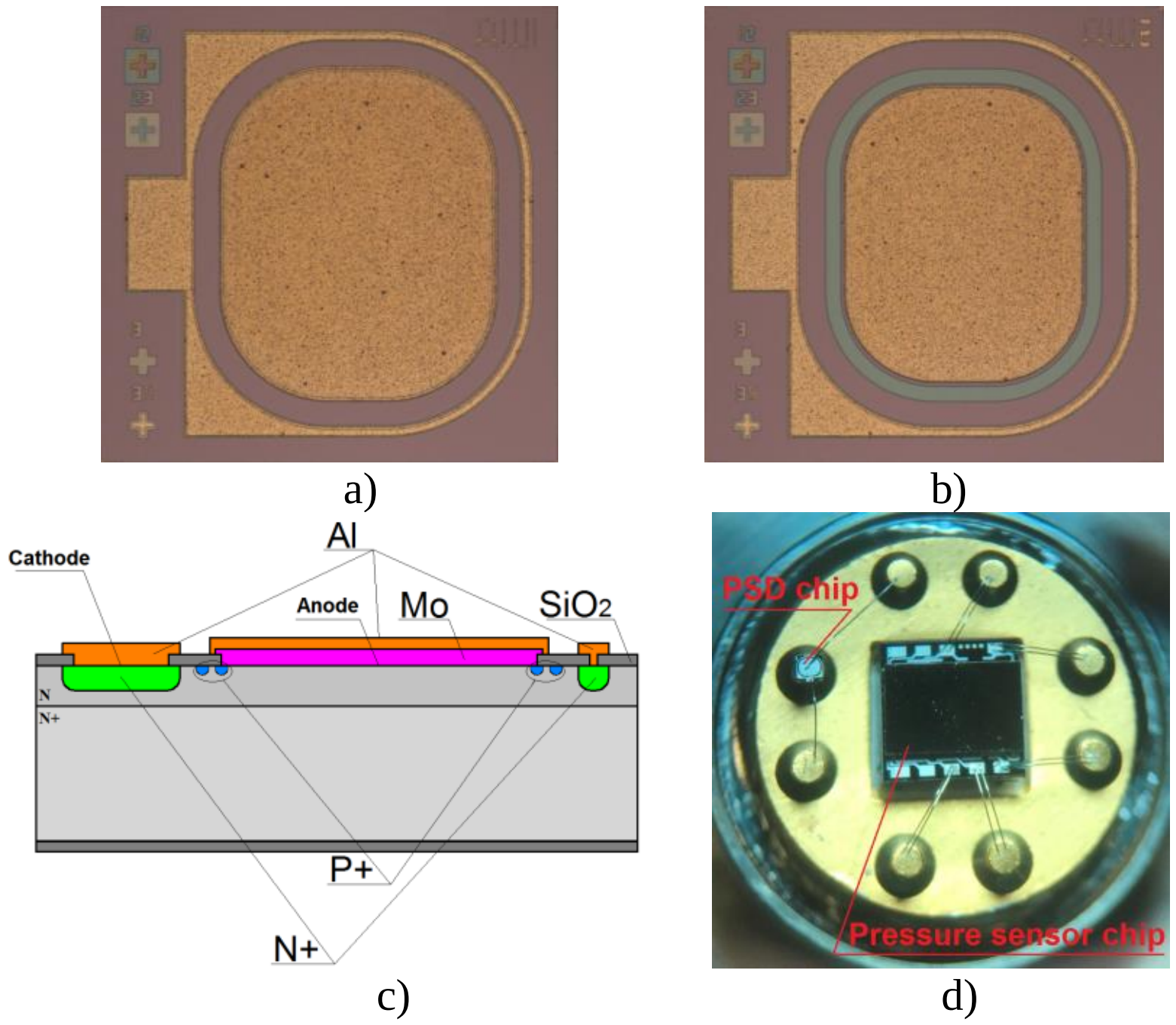

**Fig. 1.** PSD chip: a) view of No. 1, b) view of No. 2, c) schematic section of structure with two GRs (No. 2) and barrier layer of Mo, d) placement in a single case with pressure sensor.

Annealing mode in vacuum for each of the barrier metals was found [20, 64]. After that process two types of metal have the same low leakage currents ($I_L$ < 5 μA at $U_R$ = 20 V) and breakdown voltages ($U_{BR}$ > 70 V). Annealing for Al at T = 480 ºC and for Mo at T = 510 ºC during t = 15 min and P = $10^{-3}$ Pa was done. Fig. 2 shows the I-V characteristic of forward bias for wide range of supply current up to 1 A. The construction of PSD chip (at $I_F$ = 1 mA) with Mo barrier metal demonstrated lower value of forward voltage $U_{F\,Mo}$ = (208 ± 6) mV than in the case of Al barrier metal $U_{F\,Al}$ = (246 ± 6) mV despite the factually identical values of work function for Al ($\varphi_{B0\,Al}$ = 4.18 eV) and Mo ($\varphi_{B0\,Mo}$ = 4.21 eV).

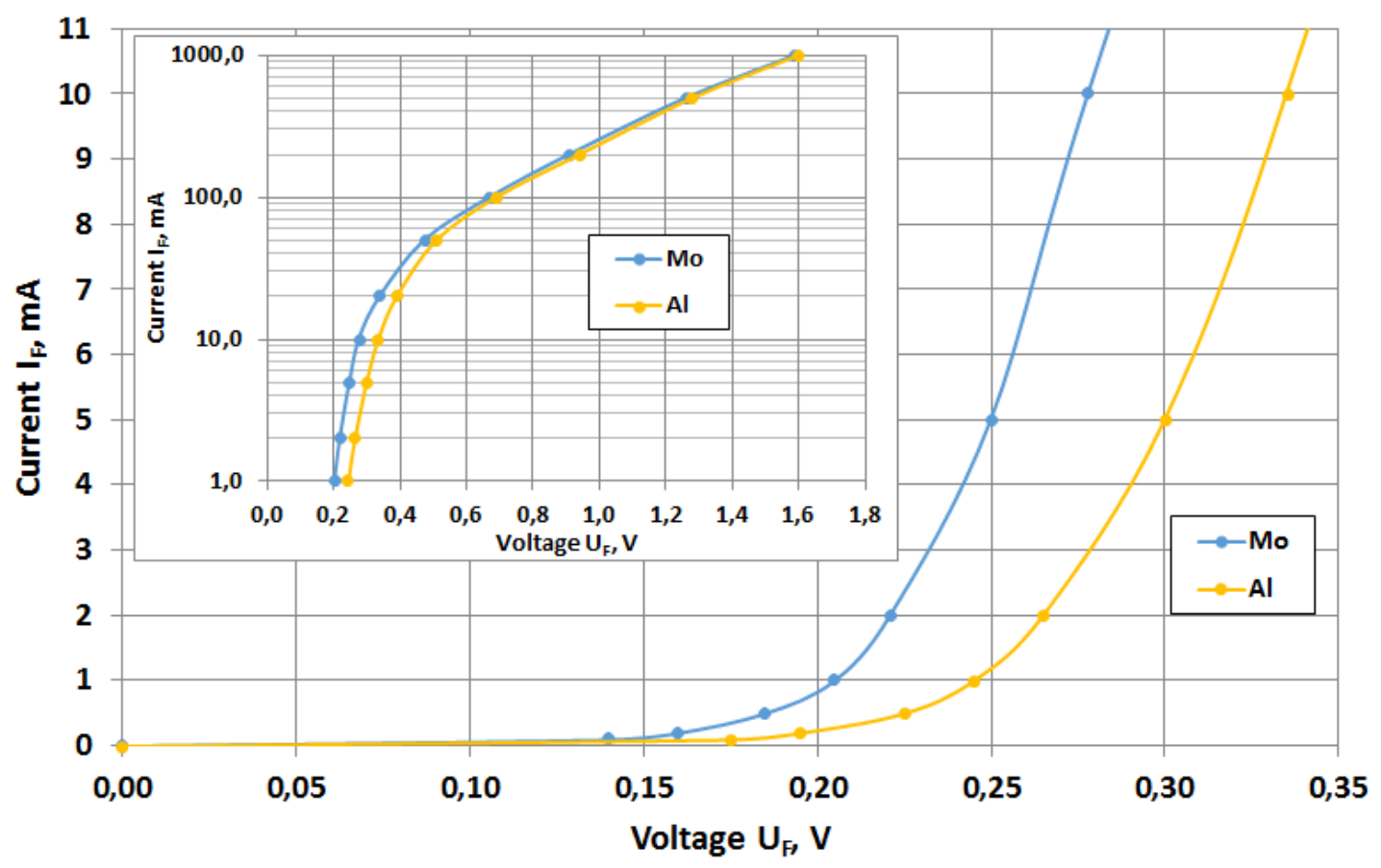

**Fig. 2.** I-V characteristic of forward branch for PSD chip with two types of barrier metal (Al and Mo).

The PSD chip No. 1 with only one GR achieved the breakdown voltage $U_{BR}$ > 70 V for reverse branch of I-V characteristics. The first probe of chip creation demonstrated similar values for the PSD chip No. 2 with two GRs. The reason for no effect of double GRs is the initial uncorrected choice of distance between GRs by PL. The gap between GRs by PL was reduced from 11 to 8 μm. It was necessary to take into account the lateral diffusion of $p^+$ -type regions (≈ 1.1 μm), $SiO_2$ etching wedge (≈ 0.2 μm) after PL and potential SCR propagation of RGs at $U_R$ ≈ 60 V (≈ 2.7 μm) [60]. The PSD chip No. 2 structure was analyzed by TCAD software (Fig. 3). Samples of the corrected PSD chip No. 2 demonstrated the achievement of breakdown voltage increase $U_{BR}$ > 85 V. The PSD chip No. 2 with Mo barrier metal is used for further studies of temperature sensor.

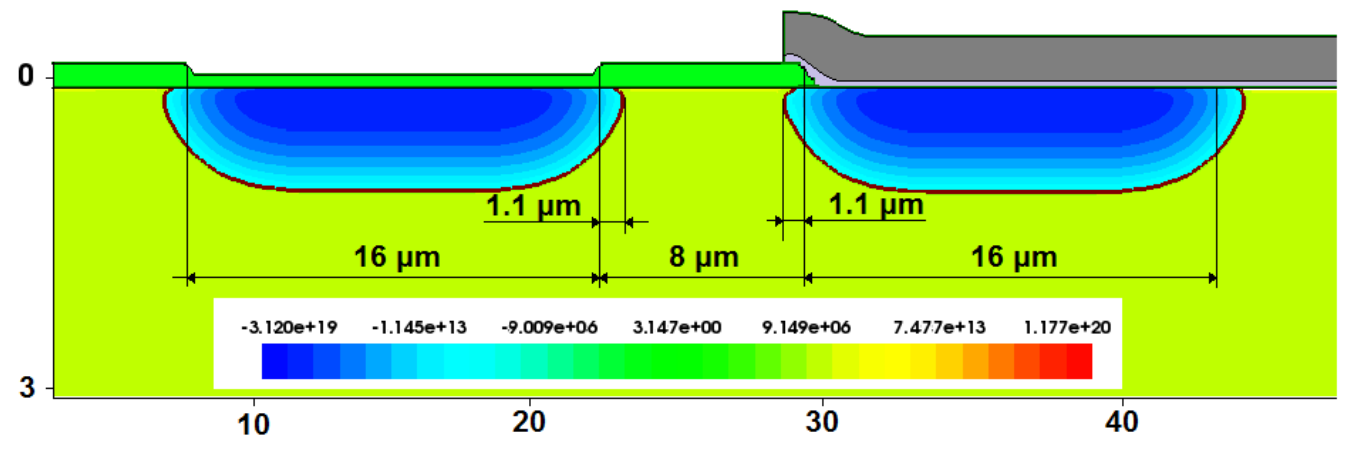

**Fig. 3.** Chip structure map (No. 2) with doping levels and distances in the GR region.

Group studies for PSD chip No. 2 as a temperature sensor were carried out using the measuring complex National Instruments PXI-1044 and thermal chamber Espec MC-811 (Fig. 4) with temperature imbalance over the full volume of less than 0.2 ºC at temperature change rate $V_T$ = 0.6 ºC/min. All measurements are taken in volume of dry air. Fig. 5 shows the temperature sensor dependences of forward voltage change on temperature (color bar - average values, dark area - spread) for two nominal supply currents $I_F$ = 1 and 10 mA. Additionally, this research has a wider range (from -65 to 165 ºC) than the operating range (from -65 to 85 ºC) for using with pressure sensor. The small jump of characteristics at temperature point of T ≈ 60 ºC is caused by the shutdown of thermal chamber refrigerator. Fig. 6 shows the dependences of nominal value and linearity error for TC (at $I_F$ = 1 mA) on various temperature ranges for measurement, where the low temperature limit remains constant ($T_{low}$ = -65 ºC) and high temperature limit $T_{high}$ varies from 75 to 165 ºC with temperature step $\Delta T_{step}$ = 10 ºC.

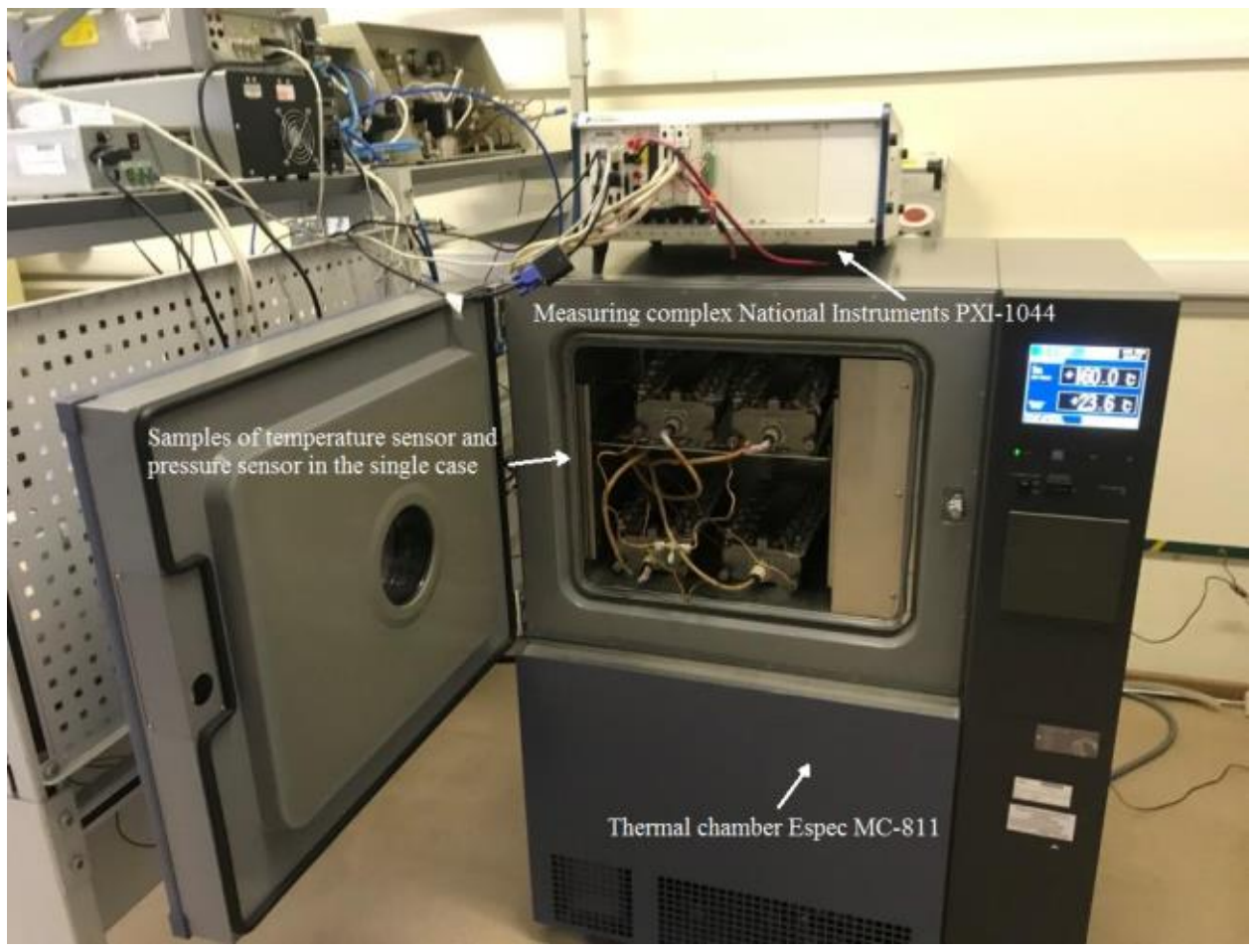

**Fig. 4.** Measuring stand for temperature sensors in a single case with pressure sensors.

The temperature sensor has average values $TK_{1mA}$ = (-1.640 ± 0.015) mV/ºC (or (-7786 ± 71) ppm/ºC) and linearity error dTK = 0.3% in the operating temperature range of pressure sensor (from -65 to 85 ºC).

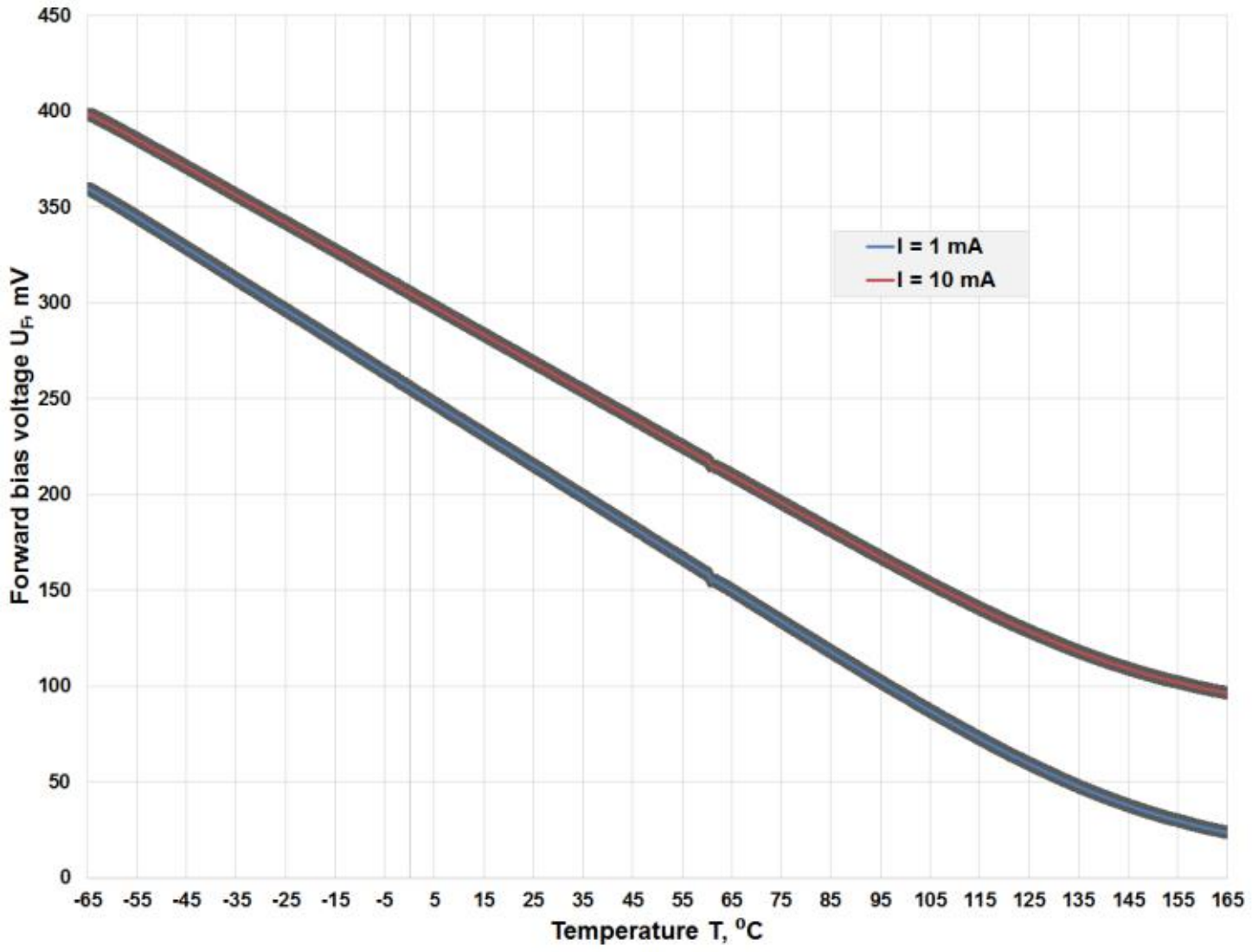

**Fig. 5.** Dependence of forward voltage of PSD chip (No. 2 with Mo) on temperature in the range from -65 to +165 ºC.

The high limit of temperature range for this sensor can be increased up to 115 ºC without significant changes in parameters. The sharp changes of linear characteristic at elevated temperatures T > 115 ºC are presented in Fig. 5 and 6: the nominal value decreases and the linearity error increases with the changing of the sign. The average TC values decrease by 9.5% ($TC_{10mA}$ = (-1.450 ± 0.013) mV/ºC) with similar linearity error dTC < 0.4% after increasing of supply current from $I_F$ = 1 mA to $I_F$ = 10 mA in the temperature range from -65 to 115 ºC. So the PSD chip can be used as a temperature sensor in wider temperature range than the operating ones due to improved reverse bias I-V characteristic.

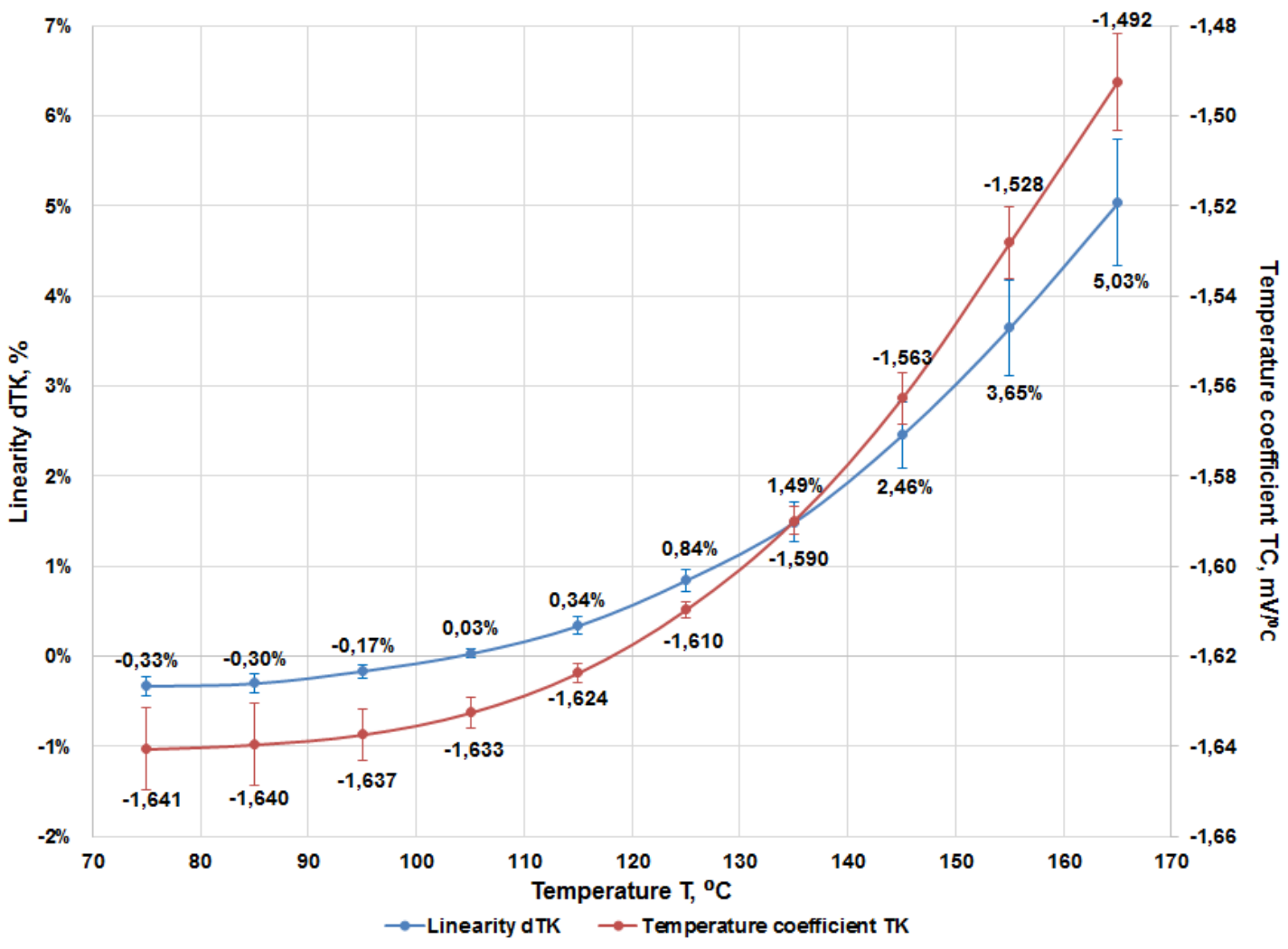

**Fig. 6.** Dependence of value and linearity error of TC on chose of the upper temperature limit.

Additional parametric characteristics of PSD chip under operating conditions by pressure sensor are presented in Table I. More than 100 samples took part in statistical data (except for less than 4% samples with a critical defect identified during testing at $T_{room}$). The temperature sensor analysis of error for forward bias voltage (supply current $I_F$ = 1 mA) was carried out according to:

- The temperature hysteresis dTH for two operating temperature sub ranges from -65 to 10 ºC and from 10 to 85 ºC (Fig.7a).
- The long-term stability $dU_{st}$ for 9 hours at T = 30 ºC.
- The effect of thermal cycling influence $dU_c$ after 5 cycles during 22 hours in temperature range from -65 to +165 ºC.
- The effect of all-round compression influence $dU_p$ by pressure $P_{com}$ = 10 MPa (Fig. 7b).

**Table 1**

Parameters of PSD chip as temperature sensor

| | Parameters | Units | Conditions | Value |
|---|---|---|---|---|
| The forward branch of I–V characteristic ($I_F$ = 1 mA) | | | | |
| | Forward bias voltage $U_F$ | mV | +25 °C | 208 ± 6 |
| | TC | mV/°C | -65…+85 °C | -1,635 ± 0,015 |
| | | ppm/°C | | 7786 ± 71 |
| Error | Linearity dTC | °C | | 0,31 ± 0,22 |
| | | % | | 0,21 ± 0,15 |
| | Hysteresis dTH | °C | -65…+10 °C | 0,07 ± 0,03 |
| | | | +10…+85 °C | 0,17 ± 0,12 |
| | | % | -65…+10 °C | 0,05 ± 0,02 |
| | | | +10…+85 °C | 0,11 ± 0,08 |
| | Long-term stability $dU_{st}$ | % | 9 hours, 30 °C | -0,03 ± 0,01 |
| | Influence of temperature cycles $dU_c$ | mV | 5 cycles during 22 hours in the range from -65 to +165 ºC | 0,58 ± 0,24 |
| | | % | | 0,23 ± 0,10 |
| | All-round compression by pressure $dU_P$ | % | 10 MPa | 1,40 ± 0,39 |
| The reverse branch of I–V characteristic ($U_R$ = 20 V) | | | | |
| | Leakage current $I_L$ | μA | +25 °C | 2,8 ± 1,2 |
| | | | +85 °C | 99,8 ± 24,4 |

The last type of the test of the effect of all-round compression influence should be noted separately. These studies are based on the application and their method includes mass pumping of compressed dry air into a limited volume (about 6000 $cm^3$). The PSD chip responds to all-round pressure and, additionally, dynamically changing temperature by external air, which clearly presented in the sharp dependences in Fig. 7b. So this temperature effect is also the reason for rather high error $dU_p$ < 1.8%. The studies of PSD chip for reverse bias of I-V characteristic were also done. The dependencies of leakage currents $I_L$ (at $U_R$ = 20 V) on rising temperature from 25 to 85 ºC are measured and showed in Fig. 7c for 10 standard samples. The values of leakage currents less than $I_L$ < 130 μA at $U_R$ = 20 V and T = 85 ºC give an opportunity for temperature sensor works with low TC error.

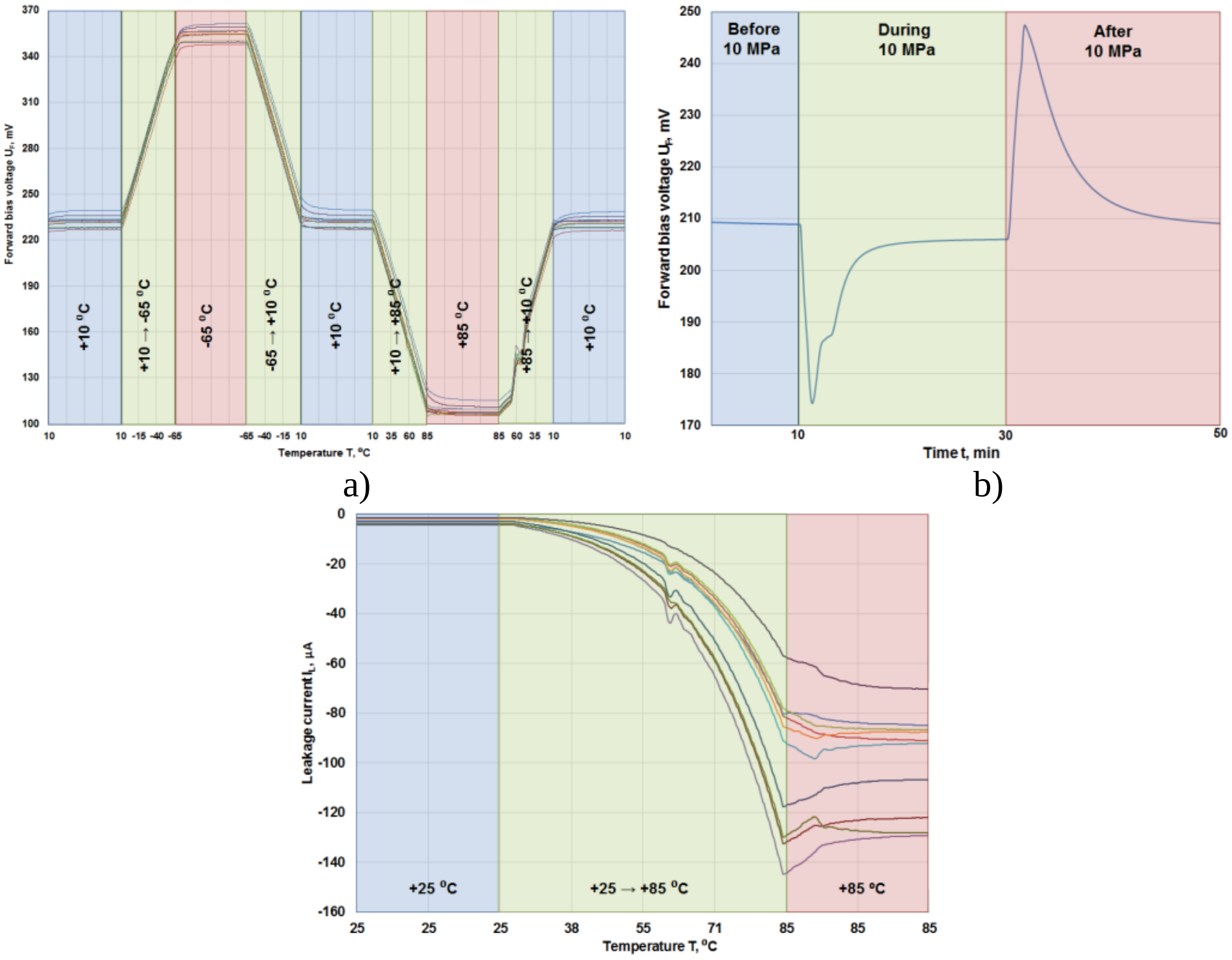

a) b)

c)

**Fig. 7.** Parametric characteristics of PSD chip for: a) temperature hysteresis of TC (forward bias voltage), b) the effect of all-round compression $dU_P$ by pressure P = 10 MPa (forward bias voltage), c) leakage current $I_L$ for reverse voltage $U_R$ = 20 V with temperature increase from 25 to 85 ºC.

## 4. Conclusion

The temperature sensor for providing of input data for temperature compensation of pressure sensor errors by ASIC and for direct temperature measurement has been developed. This chip is located at the distance of less than 1.5 mm from pressure sensor chip in the single case. The temperature is used in operating conditions of pressure sensor. The small-sized temperature sensor (0.8x0.8x0.4 $mm^3$) in the form of PSD chip based on the physical properties of Schottky barrier between metal of Mo and semiconductor Si n-type conductivity ($N_D = 3 \cdot 10^{15}$ $cm^{-3}$). Low values of forward bias voltage $U_F$ = (208 ± 6) mV at $I_F$ = 1 mA are achieved by the optimal mode of metal annealing in vacuum. The breakdown voltage is $U_{BR} > 85$ V and the leakage current $I_L < 5$ μA at T = 25 ºC and $I_L < 130$ μA at T = 85 ºC (at $U_R = 20$ V) for PSD chip. These parameters were achieved by the structure of two GRs $p^+$- type conductivity, which located at the sufficient distance to intersect their SCR. The SCRs combine at the moment close to a single breakdown each GR. The PSD chip has temperature coefficient TC = (-1.635 ± 0.015) mV/ºC (or (-7786 ± 71) ppm/ºC) with low linearity error dTC < 0.4% and temperature hysteresis dTH < 0.3% in the operating temperature range from -65 to 85 ºC and supply current $I_F = 1$ mA. Additional studies for higher temperatures demonstrated possibility of PSD chip functioning for wider range from -65 to 115 ºC without significant deviations. The increase of supply current from 1 mA to 10 mA reduces TC by 9.5% ($TC_{10mA}$ = (-1.450 ± 0.013) mV/ºC). The error under the all-round compression influence by pressure $P_{com} = 10$ MPa has high boundaries of $dU_p < 1.8\%$ because this parameter contains two parts: pressure and additional temperature change in the measuring equipment. Also, the PSD chip showed low error for long-term stability for 9 hours $dU_{st} < 0.05\%$ at T = 30 ºC and after thermal cycling $dU_c < 0.4\%$ (5 cycles during 22 hours in the range from -65 to +165 ºC). The developed temperature sensor in the form of PSD chip with small size, low power consumption, high breakdown voltage and high linear TC can be used in many of the previously mentioned industries or together with elements other than pressure sensors.

**Acknowledgment**

The studies were financially supported by Dukhov Automatics Research Institute.

## Biographies

**Basov Mikhail** graduated in 2012 from National Nuclear Research University "Moscow Engineering Physics Institute" (department of Nano- and Microelectronics). Starting from 2010 he works as R&D engineer on advanced MEMS pressure sensors and microelectronic temperature sensor. His Ph.D. thesis is titled "High-Sensitivity Pressure Sensor Chip Utilizing Bipolar-Junction Transistors". He won the award of state atomic energy corporation Rosatom with project "Ultra-High Sensitive Pressure Sensor with Increased Strength" in 2019.